\documentclass[
    ,final            
  ]
  {aipproc}

\layoutstyle{8x11single}

\def\be{\begin{equation}}
\def\ee{\end{equation}}
\def\bea{\begin{eqnarray}}
\def\eea{\end{eqnarray}}
\def\lb{\label}
\usepackage{amssymb}
\begin{document}

\title{Dark Energy: Beyond General Relativity?}

\classification{04.62.+v, 98.80.Cq}
\keywords      {Dak Energy, gravitation}

\author{David Polarski}{
  address={Laboratoire de Physique Th\'eorique et Astroparticules, UMR 5207 CNRS\\ 
Universit\'e de Montpellier II, F 34095 Montpellier cedex 05, France}
}

\begin{abstract}
The late-time accelerated expansion of the universe is a major challenge 
for cosmology. It may well be that a solution to this problem will require 
a theory of gravitation beyond General Relativity. It is emphasized that 
precision cosmology will strongly constrain the possibilities by using 
observational data probing the background as well as the inhomogeneities.  
\end{abstract}

\maketitle


\section{Introduction}
There is growing \emph{observational} evidence for a late time accelerated expansion 
of our universe \cite{Perl}. Such an expansion constitutes a radical departure from 
concentional cosmology but it is reassuring that a consistent picture of our universe 
seems to emerge from all the data \cite{BOPS99}.
This accelerated expansion might be due to a new component with sufficiently negative 
pressure, coined Dark Energy. The simplest candidate of this kind is a positive 
cosmological constant $\Lambda$ whose pressure satisfies $p_{\Lambda}=-\rho_{\Lambda}$. 
As well-known, its interpretation as the vacuum energy is problemetic because of its 
exceeding smallness (see e.g. \cite{SS00} for recent comprehensive reviews).
One can also introduce phenomenologically a smooth component 
with constant equation of state and sufficiently negative pressure typically satisfying 
$w_{DE}\equiv \frac{p_{DE}}{\rho_{DE}}<-0.5$ \cite{CP01}.   
Observations could force us to consider a smooth component with	a varying equation 
of state. The most prominent candidate in this respect is a minimally coupled 
scalar field, Quintessence models \cite{RP88}. 
A further constraint appears if observations force us to consider DE models of the 
phantom type, i.e. satisfying $w_{DE}<-1$ \cite{C02} as observations seem to suggest 
on small redshifts $z\lesssim 0.5$  \cite{ASSS04}. The most 
striking consequence in that case is that quintessence models are ruled out.
  
However, the late-time acceleration  might as well be caused by a change in the 
theory of gravitation, no longer described by General Relativity \cite{GG}. 
Ineterest for such candidates is increased by their ability to produce DE of the 
phantom type. We will review scalar-tensor DE models  \cite{BEPS00}, \cite{T02} 
as a promising candidate belonging to this class of models . 
It is emphasized that observations probing the background \emph{and} the 
inhomogeneities will establish whether such attempts provide a consistent 
description of our universe. 

\section{Two basic Dark Energy  models inside General Relativity}

\subsection{Constant equation of state DE models}
The simplest DE models are those in the framework of GR containing some unknown component with negative equation 
of state parameter $w_{DE}\equiv \frac{p_{DE}}{\rho_{DE}}$.    
Our starting point are the well-known Friedmann equations for a homogeneous and isotropic 
universe,
\bea
\biggl(\frac{\dot a}{a}\biggr)^2 + \frac{k}{a^2} &=& \sum_{i} \frac{8\pi G}{3} ~\rho_i~,\lb{FR1}\\
\frac{\ddot a}{a} &=& -\frac{4\pi G}{3} \sum_{i} (\rho_i + 3 p_i)~.\lb{FR2}
\eea
In  the equations above, the different components labelled $i$, are all isotropic perfect fluids 
while a dot stands for a derivative with respect to (cosmic) time $t$. 
We note from (\ref{FR2}) that a necessary, but not yet sufficient, condition that a component $i$ 
induces an accelerated expansion is given by  
\be
\rho_i + 3 p_i<0~.
\ee
We concentrate on equations of state of the form $\rho_i=w_i~p_i$, with constant $w_i$ and we 
specialize to a universe containing dustlike matter and some DE component. 
From (\ref{FR2}), it is not hard to deduce that our universe is presently accelerating provided 
\be
w_{DE}<-\frac{1}{3} \left( 1 + \frac{\Omega_{m,0}}{\Omega_{DE,0}} \right)~,
\ee
and in particular for a flat universe
\be
w_{DE}<-\frac{1}{3} \Omega_{DE,0}^{-1}~.
\ee
For instance, for $\Omega_{m,0}=0.3,~\Omega_{DE,0}=0.7$, $w_{DE}<-0.47$ is required. Therefore, present 
experimental evidence yielding $\Omega_{m,0}\sim 0.3$, combined with the location of the first 
acoustic peak in the CMB anisotropy first detected by balloon experiments and confirmed by the recent 
CMB data released by WMAP \cite{WMAP03} suggesting a 
nearly flat universe, imply that our universe would be presently accelerating for a wide range of 
constant values, roughly $w_{DE}<-0.5$. 
Introducing the dimensionless quantity $x\equiv \frac{a}{a_0}$, accelerated expansion starts at $x_{a}$ 
given by 
\be
x_{a}^{-3 |w_{DE}|}= (-1+ 3|w_{DE}|) ~\frac{\Omega_{DE,0}}{\Omega_{m,0}}~,\lb{xa}
\ee
which corresponds to redshifts 
\be
z_a = (-1+ 3|w_{DE}|)^{\frac{1}{3|w_{DE}|}} ~\left( \frac{\Omega_{DE,0}}{\Omega_{m,0}} \right)^{\frac{1}{3|w_{DE}|}} 
- 1~.\lb{za} 
\ee
For $-1<w_{DE}\to -1/3,~z_a$ is shifted towards smaller redshifts, for 
$(\Omega_{m,0},~\Omega_{DE,0})=(0.3,~0.7)$, we have $z_a=0.671$ for a constant $\Lambda$-term, and 
$z_a=0.414$ when $w_{DE}=-0.6$. The fact that $z_a$ is so close to zero, is the cosmic coincidence problem.
Finally, these models can yield a significant increase, depending on $w_{DE},\Omega_{m,0},\Omega_{DE,0}$, of 
the age of the universe for given Hubble parameter $H_0$ compared to an Einstein-de Sitter universe, i.e. 
a flat universe with $\Omega_{m,0}=1$ \cite{CP01}.  

\subsection{Quintessence}
The DE component could well be a time dependent minimally coupled scalar field $\phi(t)$ called Quintessence. 
This possibility is clearly inspired by the inflatonary paradigm in which a scalar field  is so successful 
in implementing the inflationary stage. Such a scalar field can be considered as a perfect fluid with
\be
\rho_{\phi} = \frac{1}{2} {\dot \phi}^2 + V(\phi) \qquad \qquad \qquad p_{\phi} =  \frac{1}{2} {\dot \phi}^2 - V(\phi)
\ee
and therefore the equation of state parameter $w_{\phi}$ is given by 
\be
w_{\phi} =  \frac{{\dot \phi}^2 - 2 V(\phi)}{{\dot \phi}^2 + 2 V(\phi)} 
\ee
For $\rho_{\phi}\ge 0$, the equation of state must satisfy 
\be
w_{\phi} \ge -1~,\lb{P}
\ee
in other words $\phi$ {\it cannot be of the Phantom type}.
It is possible to have scaling solutions with $\rho_{\phi}\propto x^m,~m=-3(1+w_{\phi})$=constant. However 
this requires a very particular potential $V(\phi)$ for which, 
\be
V(\phi) = \frac{1-w}{1+w}~\frac{\dot \phi^2}{2}~.
\ee
Hence, the most natural thing for Quintessence is to have a time varying equation of state, however one 
that satisfies the condition ($\ref{P}$). 

The two models reviewed in this section, though very different, are inside the framework of General 
Relativity (GR). A further dramatical departure is to modify the laws of gravity and to consider DE models 
outside the framework of GR. We will consider in details the example of scalar-tensor DE models. 

\section{Scalar-tensor Dark Energy models}

\subsection{Full Reconstruction}
As we have seen, in quintessence models the equation of state parameter $w_{\phi}$ 
must satisfy $w_{\phi}\ge -1$. The following inequality must be satisfied 
\begin{equation}
\frac{dH^2(z)}{dz}\ge 3 \Omega_{m,0} H_0^2 (1+z)^2~.\label{ineq}
\end{equation}
Note that inequality (\ref{ineq}) applies only to spatially flat universes, a full analysis 
should relax the flatness prior \cite{PR05}. 
It is not clear from the existing data whether~(\ref{ineq}) is satisfied and actually 
the analysis of the most recent SN data supports a varying equation of state which is of the 
Phantom type, i.e. with $w<-1$, on very small redshifts $0\le z\lesssim 0.5$. If  
confirmed, a striking consequence is that Quintessence models are ruled out.  
It is therefore important to consider a more general class of models, like scalar-tensor 
(ST) models, where the inequality~(\ref{ineq}) is no longer compulsory. 
Further ST theories are interesting to consider as they arise naturally from more fundamental 
theories like $M$-theory.

We now review the reconstruction program to scalar-tensor DE models \cite{BEPS00}.  
We consider the following Lagrangian density in the Jordan frame (JF)
\begin{equation}
L={1\over 2} \Bigl (F(\Phi)~R -
g^{\mu\nu}\partial_{\mu}\Phi\partial_{\nu}
\Phi \Bigr) - U(\Phi) + L_m(g_{\mu\nu})~,
\label{L}
\end{equation}
where $L_m$ describes dustlike matter and $F(\Phi)>0$. The Lagrangian as it is written in (\ref{L}) 
can describe consistently models with a positive Brans-Dicke parameter 
$\omega_{BD}=F/(dF/d\Phi)^2>0$. If $\omega_{BD}<0$, we must use the parametrization with $Z=-1$ where 
$Z$ is the function in front of the kinetic term, the Lagrangian (\ref{L}) corresponds to the $Z=1$ 
parametrization. We will write all our equations using (\ref{L}), i.e. in the JF using the 
$Z=1$ parametrization. We do not introduce any direct coupling between $\Phi$ and $L_m$ so that, 
in particular, fundamental constants do not change with time. 
We can proceed in a way analogous to the Quintessence case, the main difference being that we have 
now to reconstruct two unknown functions instead of one.
Again, specializing to a flat FRW universe, the corresponding (modified) Friedmann equations read 
\bea
3FH^2 &=& \rho_m + {\dot \Phi^2\over 2} + U - 3H {\dot F}~,\lb{F1}\\
-2 F {\dot H} &=& \rho_m + \dot \Phi^2 + {\ddot F} - H {\dot F}~.\lb{F2}
\eea
As was the case in GR, the evolution equation for the scalar field $\Phi$ is contained 
in the two Friedmann equations above. These background equations (\ref{F1},\ref{F2}) can be 
combined to give the following master equation for $F(z)$:
\begin{eqnarray}
F'' &+& \left[(\ln H)' - \frac{4}{1+z}\right]~F' +
\left[\frac{6}{(1+z)^2} - \frac{2(\ln H)'}{1+z}\right]~F
\nonumber\\
&=& \frac{2U}{(1+z)^2 H^2} + 3~(1+z)
\left(\frac{H_0}{H}\right)^2 F_0~\Omega_{m,0}~,\lb{F}
\end{eqnarray}
where a prime denotes a derivative with respect to $z$.

In this theory, the effective value of Newton's gravitational constant $G_N$ is given by 
\be
G_N=\frac{1}{8\pi F}~.\lb{GN} 
\ee
It is natural to use its present value $G_{N,0}\equiv \frac{1}{8\pi F_0}$ 
in the definition of the critical density $\rho_{\rm crit}$.
However $G_N$ does not have the same physical meaning as in GR, it is no longer the coupling constant 
for the gravitational attraction between two point masses. For a massless dilaton, the effective 
gravitational constant between two test masses is given by
\begin{equation}
G_{\rm eff} = {1\over 8\pi F}
\left({2F+4(dF/d\Phi)^2\over 2F+3(dF/d\Phi)^2}\right)~.\lb{Geff}
\end{equation}
In our case, the dilaton is massive but (\ref{Geff}) will still hold for physical scales $R$ such that
\begin{equation}
R^{-2} \gg \max \left(\left|{d^2U\over d\Phi^2}\right|, H^2,
H^2 \left|{d^2F \over d\Phi^2}\right|\right)~.\lb{R}
\end{equation}

The most recent solar system measurements~\cite{BIT03} imply very stringent constraints 
on the Brans-Dicke parameter {\it today}
\begin{equation}
\omega_{BD,0}=F_0~ \left( \frac{d\Phi}{dF}\right)_0^2 > 4\times 10^4~.\label{BD0}
\end{equation}
As a consequence, $G_{N,0}$ and $G_{{\rm eff},0}$ coincide with better than
$1.25\times 10^{-5}$ accuracy. On the other hand, the difference between
$G_N$ and $G_{\rm eff}$ could be larger at higher redshifts. Our interest for scalar-tensor 
theories of gravity is in the context of DE models. 
So our theory should satisfy the following requirements as any realistic DE model. 
First, the DE term should dominate today the energy density of the universe and satisfy 
\be
\Omega_{DE,0}\sim 0.7\sim 2 \Omega_{m,0}~.
\ee
If our model describes a universe whose expansion is presently accelerated, then it must satisfy 
\be
U_0>(\rho_m+2\dot\Phi^2+3\ddot F+3H\dot F)_0
\ee
Finally it is important that DE remains essentially unclustered at scales up to $R\sim 10h^{-1}(1+z)^{-1}$ Mpc, 
though as we will see some limited clustering will arise on these scales. To achieve this, it is 
sufficient to assume that the inequality (\ref{R}) is satisfied for all scales of interest.

Like in GR, we start with the determination of $H(z)$ from $D_L(z)$ using 
\be
{1\over H(z)} = \left(\frac{D_L(z)}{1+z} \right)'~.\label{DL}
\ee
However, we need to recover the two functions $F(z)$ and $U(z)$, so the substitution of $H(z)$ 
in (\ref{F}) is no longer sufficient. For a complete reconstruction we must use a new equation 
based on independent observations. It is provided by $\delta_m(z)$ data, measurement of the 
linear dustlike matter density fluctuations. We can expect in the near future accurate $D_L(z)$ 
and $\delta_m(z)$ data. 

We consider the perturbation equations in the longitudinal gauge 
\be
ds^2= -(1 + 2 \phi) dt^2 + a^2 (1 - 2\psi) d{\bf x}^2~, 
\ee
(see \cite{BEPS00} for details). The idea is that, in the short wavelength limit, the leading terms 
are either those containing $k^2$, or those with $\delta_m$.
Then the following equation is obtained 
\begin{equation}
\delta \Phi \simeq (\phi - 2\psi)~{dF\over d\Phi} \simeq
- \phi~{F~dF/d\Phi\over F+2(dF/d\Phi)^2}~.
\label{dPhi}
\end{equation}
Hence, unlike in GR, in ST gravity the dilaton remains partly clustered even for arbitrarily small 
scales, however this clustering is small because $\omega_{BD}$ is large. In the same short 
wavelength limit, Poisson's equation has the same form as in GR, with the important 
difference that Newton's constant $G_N$ is replaced by $G_{\rm eff}$, defined in (\ref{Geff}) above. 
Hence, the equation for the evolution of dustlike matter linear density perturbations finally leads 
to 
\be
H^2~\delta_m'' + \left(\frac{(H^2)'}{2} -
{H^2\over 1+z}\right)\delta_m'
\simeq {3\over 2} (1+z) H_0^2 {G_{\rm eff}(z)\over
G_{N,0}}~\Omega_{m,0}~\delta_m~.\lb{delz}
\ee
So, we see here a second difference in the physical meaning of $G_{\rm eff}$ and $G_N$: while it is 
$G_N$ that appears in the equation for the evolution of the perturbations in GR, in ST models 
this role is played by $G_{\rm eff}$.

Let us now sketch briefly the reconstruction itself. 
Extracting $H(z)$ (through $D_L(z)$) and $\delta_m(z)$ from observations with sufficient accuracy, 
we first reconstruct $G_{\rm eff}(z)/G_{N,0}$ analytically. Since, as follows from Eq.(\ref{BD0}), 
the quantities $G_{{\rm eff},0}$ and $G_{N,0}$ coincide with better than $0.00125\%$ accuracy, 
Eq.(\ref{delz}) taken at $z=0$ gives also the value of $\Omega_{m,0}$ with the same accuracy. 
Thus, in principle, no independent measurement of $\Omega_{m,0}$ is required, it follows from 
(\ref{delz}) taken at $z=0$

We get an equation $G_{\rm eff}(z)=p(z)$, where $p(z)$ is a given function that can be determined 
solely using observational data, which can be transformed into a nonlinear second order differential 
equation for $F(z)$ using the background equation 
\be
\Phi'^2 = - F'' - \left[(\ln H)' + \frac{2}{1+z}\right]~F' +\frac{2(\ln H)'}{1+z}~F
-3(1+z) \frac{H_0^2}{H^2} F_0~\Omega_{m,0}~.\label{Phi}
\ee
Hence $F(z)$ can be determined by solving the equation $G_{\rm eff}(z)=p(z)$ after we supply the 
initial conditions $F_0=\frac{1}{8\pi G_{N,0}}$ and $F'_0$. Actually $F'_0$ must be very close to 
zero due to the solar system constraint (\ref{BD0}). Once $F(z)$ is found, it can be substituted 
into equation (\ref{F}) to yield the potential $U(z)$ in function of redshift.
Then, using Eq.~(\ref{Phi}), $\Phi(z)$ is found by simple integration which, after inverting this 
relation, gives us $z=z(\Phi-\Phi_0)$. Finally, both unknown functions $F(\Phi)$ and 
$U(\Phi)$ are completely fixed as functions of $\Phi-\Phi_0$. Of course this reconstruction can 
only be implemented in the range probed by the data corresponding to $z\lesssim 2$. 

\subsection{Partial reconstruction}
As one does not expect to have in the very near future data referring to perturbations that are 
as accurate as the distance-luminosity data, it is interesting to try to extract as much information 
as possible using only $D_L(z)$ data. If we make assumptions on either $F(z)$ or $U(z)$, or if we 
assume some functional relation between both, it is again possible to reconstruct the theory using only 
$D_L(z)$ data. This is what we mean by reconstruction of constrained models, or partial reconstruction: 
we reconstruct models where {\it by assumption} there is effectively only one unknown function. Several 
cases have been considered in \cite{EP01} and powerful constraints can be obtained. In particular, 
it is interesting that these constraints can go beyond solar-system constraints just because they go 
back in time and probe the cosmological evolution.
To see how powerful constraints can be obtained, we consider the interesting question whether it is 
possible to have a vanishing potential $U$ while the expansion of the universe still satisfies  
\be
H^2(z) = H_0^2~[0.3~(1+z)^3 + 0.7]~.\lb{H2}
\ee
Of course in the framework of GR an expression like (\ref{H2}) lends itself to the straightforward 
interpretation of a flat $\Lambda$ dominated universe with the cosmological parameters 
\be
\Omega_{\Lambda,0} \approx 0.7~,~\qquad\qquad  \Omega_{m,0} \approx 0.3~.\lb{OmegaValues}
\ee
The dynamical laws of GR are encoded in the definition of these cosmological parameters through the Friedmann 
equations. Now, we assume we have the same kinematics (expansion) but with modified dynamical laws as we 
deal with a modified gravity theory. 
When $U=0$ equation (\ref{F}) can be solved numerically and one can show that $F(z)$ will vanish at 
most at $z_{max}\approx 0.66$ if $F(z)$ evolves according to (\ref{F}). Therefore, the predicted vanishing 
of $F(z)$ is well inside the range for which $H(z)$ is determined with sufficient accuracy. Hence, 
for $H(z)$ obeying (\ref{H2}) up to $z\approx 0.66$, these models with vanishing potential $U$ are 
excluded. 

If observations force us to consider models that can have DE of the phantom type then quintessence 
models, i;e. models inside General Relativity with a minimally coupled scalar (quintessence) field, 
are ruled out and is led to consider models outside GR.like scalar-tensor DE models that we have 
been discussing here. Many problems appearing in these models are representative of all DE models 
outside GR. Clearly discrimination between all the models will accurate data probing the background 
{\it and} the inhomogeneities: Supernovae data, Cosmic Microwave Background data, galaxy surveys, weak 
lensing data, etc. (see e.g. \cite{IUS05}).      

Recently, investigation of $f(R)$ modified gravity theories led to the surprising result \cite{APT06}  
that the cosmological history for a large class of such models is incompatible with observations 
because of the disappearance of the usual matter-dominated stage with $a\propto t^{\frac{2}{3}}$ which is 
replaced by a stage with $a\propto t^{\frac{1}{2}}$. So in this case, cosmological background constraints 
come from large $z$ behaviour and not just from SNIa data on small redshifts $z<2$!


\begin{theacknowledgments}
It is a pleasure to thank my collaborators B. Boisseau, G. Esposito-Far\`ese, A. Starobinsky.  
\end{theacknowledgments}







\end{document}